\documentclass[sigconf]{acmart}

\usepackage{algorithm}
\usepackage{algorithmic}
\usepackage{multirow}
\usepackage{threeparttable}

\AtBeginDocument{%
  \providecommand\BibTeX{{%
    \normalfont B\kern-0.5em{\scshape i\kern-0.25em b}\kern-0.8em\TeX}}}

\setcopyright{acmcopyright}
\copyrightyear{2020} 
\acmYear{2020} 
\setcopyright{acmcopyright}\acmConference[MM '20]{Proceedings of the 28th ACM International Conference on Multimedia}{October   12--16, 2020}{Seattle, WA, USA}
\acmBooktitle{Proceedings of the 28th ACM International Conference on Multimedia (MM '20), October 12--16, 2020, Seattle, WA, USA}
\acmPrice{15.00}
\acmDOI{10.1145/3394171.3413623}
\acmISBN{978-1-4503-7988-5/20/10}

\begin{document}

\fancyhead{} 

\title{DualLip: A System for Joint Lip Reading and Generation}




\author{Weicong Chen}
\affiliation{%
 \institution{Tsinghua University}
}
\email{chenwc18@mails.tsinghua.edu.cn}

\author{Xu Tan}
\affiliation{%
 \institution{Microsoft Research Asia}
}
\email{xuta@microsoft.com}

\author{Yingce Xia}
\affiliation{%
 \institution{Microsoft Research Asia}
}
\email{yingce.xia@microsoft.com}

\author{Tao Qin}
\affiliation{%
 \institution{Microsoft Research Asia}
}
\email{taoqin@microsoft.com}

\author{Yu Wang}
\authornote{Corresponding author}
\affiliation{%
 \institution{Tsinghua University}
}
\email{yu-wang@tsinghua.edu.cn}

\author{Tie-Yan Liu}
\affiliation{%
 \institution{Microsoft Research Asia}
}
\email{tyliu@microsoft.com}


\begin{abstract}
Lip reading aims to recognize text from talking lip, while lip generation aims to synthesize talking lip according to text, which is a key component in talking face generation and is a dual task of lip reading. Both tasks require a large amount of paired lip video and text training data, and perform poorly in low-resource scenarios with limited paired training data. In this paper, we develop DualLip, a system that jointly improves lip reading and generation by leveraging the task duality and using unlabeled text and lip video data. The key ideas of the DualLip include: 1) Generate lip video from unlabeled text using a lip generation model, and use the pseudo data pairs to improve lip reading; 2) Generate text from unlabeled lip video using a lip reading model, and use the pseudo data pairs to improve lip generation. To leverage the benefit of DualLip on lip generation, we further extend DualLip to talking face generation with two additionally introduced components: lip to face generation and text to speech generation, which share the same duration for synchronization. Experiments on GRID and TCD-TIMIT datasets demonstrate the effectiveness of DualLip on improving lip reading, lip generation and talking face generation by utilizing unlabeled data, especially in low-resource scenarios. Specifically, on the GRID dataset, the lip generation model in our DualLip system trained with only 10\% paired data and 90\% unpaired data surpasses the performance of that trained with the whole paired data, and our lip reading model achieves 1.16\% character error rate and 2.71\% word error rate, outperforming the state-of-the-art models using the same amount of paired data.
\end{abstract}


\begin{teaserfigure}
  \includegraphics[width=\textwidth]{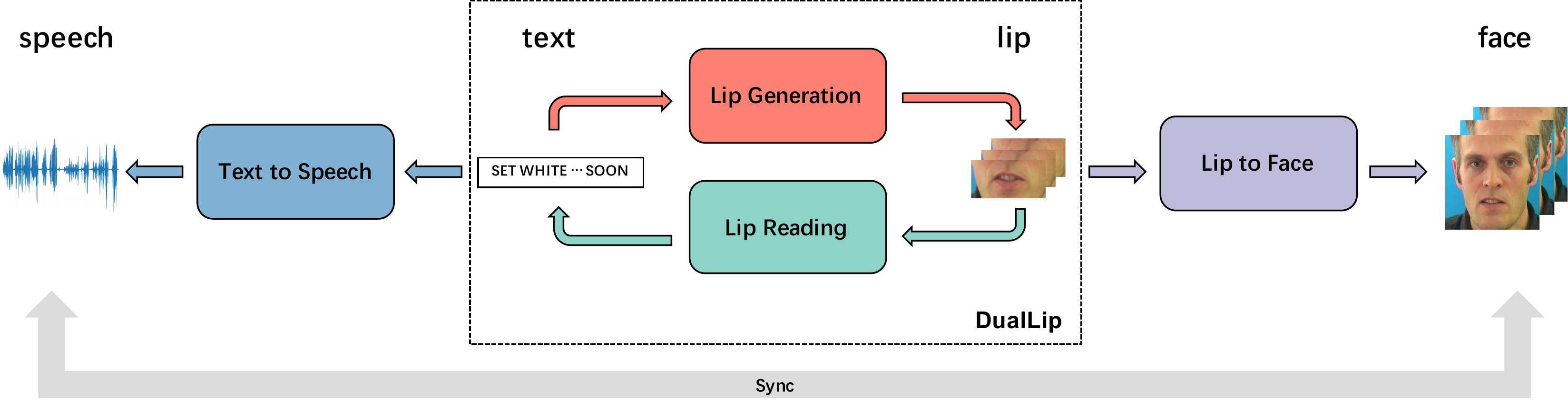}
  \caption{The framework of DualLip and text to talking face generation systems.}
  \label{fig:framework}
\end{teaserfigure}

\maketitle

\section{Introduction}

Lip reading, which aims to understand speech by visually interpreting the movements of the lips, is beneficial for deaf and hard-of-hearing people or when accompanying sound is noisy or unavailable~\cite{mcgurk1976hearing,zhu2020deep}. Lip movement generation (a.k.a. lip generation), which aims to synthesize mouth movements according to text, is an essential component of talking face generation~\cite{chung2017you,kumar2017obamanet}, which generates the video and audio of talking face from text and cover many application scenarios such as virtual assistant, avatar, etc.

Previous works on lip reading and talking face generation usually handle them separately as two independent tasks. 
Although those algorithms have made good advancements for the two tasks, they require a large amount of paired/labeled training data, which is expensive to obtain in the real world.

In this paper, we consider the task of lip generation as a dual task of lip reading and build a system named DualLip to jointly train two models for the two tasks by leveraging their task duality and unpaired/unlabeled data~\cite{he2016dual,kaneko2018cyclegan,ren2019almost}: 1) Given unlabeled text, we first use a lip generation model to synthesize lip video, and then add the pseudo paired lip video and text into the original paired data to train/improve the lip reading model. 2) Given unlabeled lip video, we first use a lip reading model to generate text, and then add the pseudo paired text and lip video into the original paired data to train/improve the lip generation model (see Section \ref{sec:dual_pipeline} to \ref{sec:lipg}). To enjoy the improvements of DualLip on lip generation, we build a complete system for talking face generation by introducing two additional components: lip to face generation to synthesize the video of talking face and text to speech generation to synthesize the audio of talking face (see Section \ref{sec:ttf}).

There are two technical challenges to build our DualLip and talking face generation systems. First, the text (input) and lip video (output) in our lip generation model are of different lengths, so we need to carefully align video frames and text characters in training and inference. Second, the two outputs (video and audio) of talking face generation system need to be synchronized to generate natural talking face.

To address the first challenge, we study two alignment settings: 1) When the duration of each character (i.e., the number of video frames corresponding to each character) is pre-given, we directly expand the text to the same length as the video according to the duration and design a lip generation model with duration as input (see Section \ref{sec:lg_dur}). 2) When the duration is not pre-given, we design a lip generation model without duration as input that leverages location sensitive attention~\cite{chorowski2015attention} to automatically learn the alignment through encoder-decoder attention between text and lip video (see Section \ref{sec:lg_nodur}).

To address the second challenge, we design a lip to face generation model that supports customized duration (see Section \ref{sec:ltf}). Meanwhile, we leverage FastSpeech~\cite{ren2019fastspeech,ren2020fastspeech}, a parallel text to speech model that also accepts customized duration as input (see Section \ref{sec:tts}). Then we set the same duration as inputs for the two models, which is either pre-given or extracted from the learned alignment of the lip generation model (see Section \ref{sec:sync}). In this way, the generated face video and speech are synchronized.

Our main contributions are as follows:

1) We design the DualLip system that leverages the task duality of lip reading and lip generation to improve both tasks by using unlabeled text and lip video, especially in low-resource scenarios where the paired training data is limited.

2) To tackle the alignment problem between input text and output lip video in lip generation, we design two models: lip generation with duration and lip generation without duration. The former achieves alignment by pre-given duration, while the latter automatically learns the alignment through attention mechanism.

3) We extend the lip generation component in DualLip and build a text to talking face generation system by introducing two additional components: lip to face generation and text to speech generation, both of which accept customized duration as inputs and thus ensure synchronization between the generated face video and audio.

4) Extensive experiments on the GRID and TCD-TIMIT datasets demonstrate the effectiveness of our systems in improving the performance of lip reading, lip generation and talking face generation. Specifically, on GRID, our lip generation model trained with only 10\% paired data and 90\% unpaired data achieves better performance than that trained with the whole paired data, and our lip reading model trained with the whole paired data and additional unpaired data outperforms the state-of-the-art models.

\begin{figure*}[htb]
  \centering
  \includegraphics[width=0.95\linewidth]{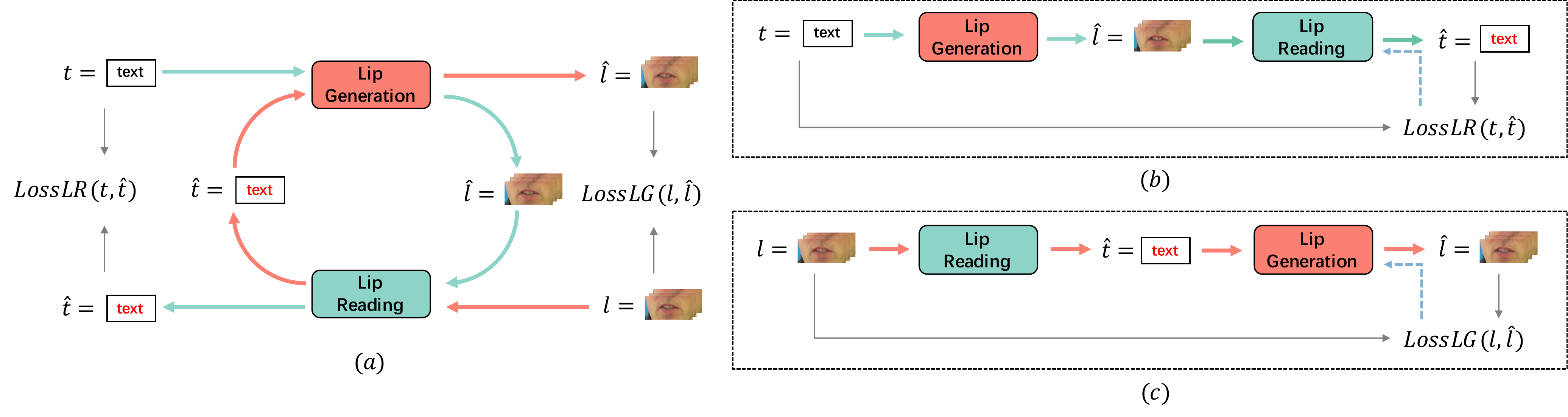}
  \caption{(a) Overview of the DualLip system. The top two straight arrows and bottom two straight arrows represent the pipelines of training the lip generation model and lip reading model under supervised training setting, respectively. For the case of unsupervised training, the green arrows represent the pipeline of training the lip reading model using unlabeled text, and the red arrows show the pipeline of training the lip generation model with unlabeled lip video. The unrolled process of unsupervised training is showed in (b) for training lip reading and (c) for training lip generation, where the dotted line indicates the direction of gradient propagation.}
  \label{fig:dual_overview}
  \Description{DualLip system overview.}
\end{figure*}

\section{Background}

\subsection{Lip Reading}
Lip reading aims to recognize spoken language using only the video of lip movements, which has many applications such as text transcription when audio is not available. Even when audio is available, lip reading can still help improve the transcript performance. Previous works on lip reading have switched from single word classification~\cite{wand2016lipreading,chung2016lip} to sentence-level prediction~\cite{Assael2016LipNet,chung2017lip,stafylakis2017combining,afouras2018deep}, and explored different models from CTC based~\citep{Assael2016LipNet,stafylakis2017combining,xu2018lcanet}, LSTM based~\cite{chung2017lip} to Transformer based~\cite{martinez2020lipreading,afouras2018deep} models to improve the performance.

Advanced models for lip reading systems rely on a large amount of paired training data to ensure accuracy. Existing datasets such as GRID~\cite{Cooke2006An} and TCD-TIMIT~\cite{Harte2015TCD} 
usually require many human efforts for data collection, while the lip reading models still prefer more paired training data. In this paper, we propose to take advantage of unpaired data by leveraging the task duality between lip generation and lip reading to help improve its performance.

\subsection{Talking Face Generation}
Talking face generation has many practical applications on virtual assistant and avatar, which can make the synthesized face more realistic.
Previous works on talking face generation usually take speech as input~\cite{chung2017you,suwajanakorn2017synthesizing,zhou2019talking}, where the input speech is usually frame-wisely synchronized with the output video. Considering text can be obtained more easily than speech, some works either take text as input to first synthesize speech and then generate talking face given synthesized speech~\cite{kumar2017obamanet,kr2019towards}, or directly take text as input to generate talking face~\cite{yu2019mining,fried2019text}. Considering the difficulty of generating the whole talking face, some works also try to first synthesize the key points or talking lip and then synthesize the full talking face given a reference face and generated talking lip~\cite{suwajanakorn2017synthesizing,kumar2017obamanet,zheng2020photorealistic}.

Talking face generation usually requires a large amount of paired data for model training and also suffers from the low-resource issue as in lip reading. In this paper, we leverage the task duality between lip reading and lip generation to help the two tasks with additional unlabeled text and lip videos. As lip generation is a key component in talking face generation, it can also benefit from the improvements of lip generation based on our proposed DualLip system.

\subsection{Task Duality}
Task duality is explored in a variety of tasks in previous works. Dual learning is proposed in~\cite{he2016dual} for neural machine translation, which leverages the task duality of two translation directions, such as English to French translation and French to English translation, with the help of additional monolingual data. The cycle consistency loss between two dual tasks is leveraged into many image and speech tasks such as image translation~\cite{zhu2017unpaired,yi2017dualgan} and voice conversion~\cite{kaneko2018cyclegan}. For the task duality across modalities, ~\cite{ren2019almost,xu2020lrspeech} improve the accuracy of automatic speech recognition and text to speech tasks with unlabeled text and speech data with dual transformation. Some works also explore the task duality between other modalities such as text and image~\cite{zhao2017dual} for image captioning, text and video~\cite{liu2019cross} for video generation. In this paper, we explore the task duality between text and lip video by leveraging the dual transformation proposed in~\cite{ren2019almost}. Unlike that in~\cite{liu2019cross}, the transformation between text and lip video is almost of no information loss, which is a good scenario to explore the dual nature of two tasks.

\begin{figure*}[ht]
  \centering
  \includegraphics[width=0.95\linewidth]{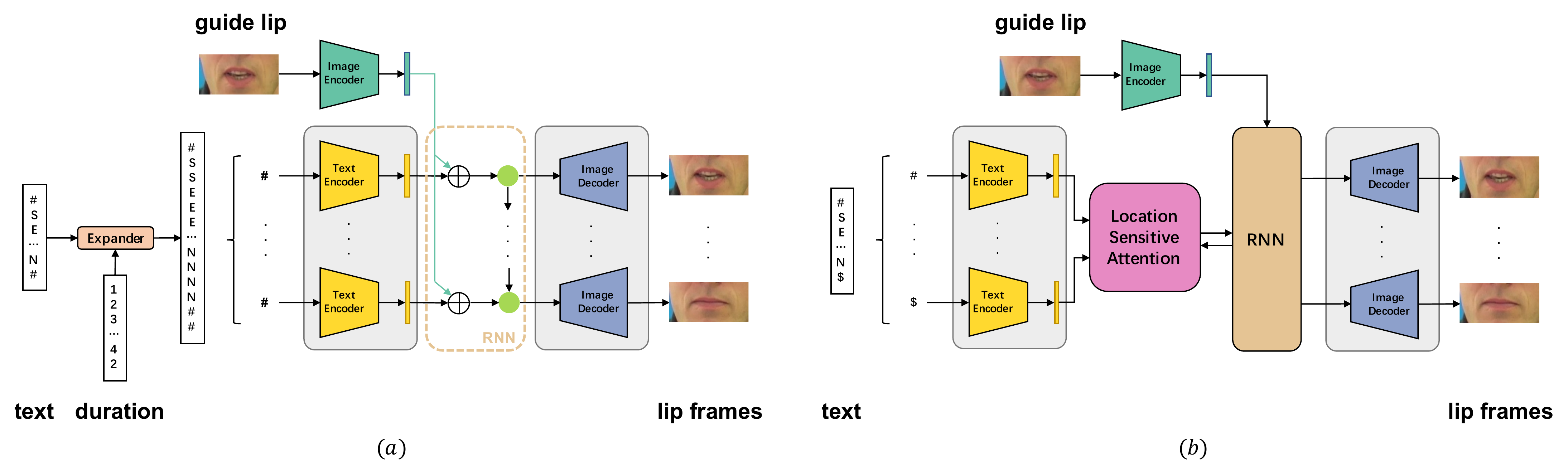}
  \caption{The model structures for lip generation model (a) w/ duration and (b) w/o duration.}
  \label{fig:lip_generation_model}
\end{figure*}

\section{Method}


\begin{algorithm}[htb]
\caption{ The DualLip pipeline.}
\label{alg:dual_lip}
\begin{algorithmic}[1] 
\REQUIRE
Paired text and lip dataset $D^p$, text-only dataset $T^u$, lip-only dataset $L^u$, lip generation model $\Theta _{lg}$, lip reading model $\Theta _{lr}$ , unsupervised loss coefficient $\alpha$.

\REPEAT
\STATE \textbf{\uppercase\expandafter{\romannumeral1}. Supervised training with text-lip data pairs}
\STATE Sample paired text and lip ($t^p$, $l^p$) from $D^p$ .
\STATE Generate lip $\hat{l}^p$ by lip generation model, and text $\hat{t}^p$ by lip reading model: \\
$\hat{l}^p = f(t^p, \Theta _{lg})$, $\hat{t}^p = f(l^p, \Theta _{lr})$
\STATE Calculate the loss for lip generation and lip reading: \\
$L^p_{lg} = LossLG(l^p, \hat{l}^p)$, $L^p_{lr} = LossLR(t^p, \hat{t}^p)$

\STATE \textbf{\uppercase\expandafter{\romannumeral2}. Unsupervised training with unpaired text and lip}
\STATE Sample unpaired text $t^u$ from $T^u$, and unpaired lip $l^u$ from $L^u$.
\STATE Generate the pseudo lip $\hat{l}$ by lip generation model, and pseudo text $\hat{t}$ with lip reading model: \\
$\hat{l} = f(t^u, \Theta _{lg})$, $\hat{t} = f(l^u, \Theta _{lr})$
\STATE Train the lip reading model with pseudo data pair ($\hat{l}$, $t^u$), and lip generation model with pseudo data pair ($\hat{t}$, $l^u$): \\
$\hat{t}^u = f(\hat{l}, \Theta _{lr})$, $\hat{l}^u = f(\hat{t}, \Theta _{lg})$
\STATE Calculate the loss for lip reading and lip generation: \\
$L^u_{lr} = LossLR(t^u, \hat{t}^u), L^u_{lg} = LossLG(l^u, \hat{l}^u)$

\STATE \textbf{\uppercase\expandafter{\romannumeral3}. Parameter update}
\STATE Calculate the total loss
\begin{equation}\nonumber
L = (L^p_{lg} + L^p_{lr}) + \alpha (L^u_{lg} + L^u_{lr})
\end{equation}
Update lip generation and lip reading model parameters with the gradient from the derivative of $L$ w.r.t $\Theta _{lg}$ and $\Theta _{lr}$ .
\UNTIL convergence

\end{algorithmic}
\end{algorithm}

\subsection{DualLip System}
\label{sec:dual_pipeline}

The overview of our proposed DualLip system is shown in Figure \ref{fig:dual_overview}, which consists of two key components: lip generation and lip reading. Dual transformation~\cite{ren2019almost} is the key idea in leveraging the task duality of lip generation and lip reading. The lip generation model transforms text to lip, while the lip reading model transforms lip to text. This way, a closed loop is formed between text and lip, which allows us to train our models on both labeled data and unlabeled data.

For supervised training with paired data, both models can be trained independently by minimizing the loss between the predicted output and the ground truth. However, under the setting of unsupervised training with unpaired data, the two models should support each other through the closed loop formed by dual transformation. Given unlabeled text sequence $t$ and unlabeled lip video sequence $l$, we first generate lip video sequence $\hat{l}$ from text sequence $t$ by the lip generation model, and then train the lip reading model on the pseudo data pair ($\hat{l}$, $t$). Similarly, we train the lip generation model on the pseudo data pair ($\hat{t}$, $l$), where $\hat{t}$ is generated by the lip reading model given $l$.

The training procedure is presented in Algorithm \ref{alg:dual_lip}, which is a two-stage procedure. In the first stage, we only conduct supervised training, aiming to let the two models learn the basic generation ability. In the second stage, we add unsupervised training into the procedure. Dual transformation is performed on the fly, where the lip generation model is trained on the newest data generated by lip reading, and vice versa, enabling the two networks to promote each other and improve together. The two stages are controlled by the unsupervised loss coefficient $\alpha$, and $\alpha = 0$ means the first stage. In the next two subsections, we introduce the model structure of lip reading and lip generation model, respectively.

\subsection{Lip Reading}
\label{sec:lipr}
Given a silent lip video sequence $l = \{ l_1, l_2, ..., l_K \}$ where $l_i$ is the $i_{th}$ lip frame, lip reading aims to predict the text $t$ that the lip is speaking. Generally, a lip reading model based on the neural network consists of two modules~\cite{afouras2018deep}: 1) a spatio-temporal visual module that extracts visual feature of each frame, and 2) a sequence processing module that maps the sequence of per-frame visual features to the corresponding character probability. The spatio-temporal visual module is a stack of 3D and 2D CNN, while the sequence processing module consists of several sequence processing layers, such as GRU, LSTM, etc.
The model configuration is presented in Section \ref{sec:model}. We adopt the connectionist temporal classification (CTC) loss~\cite{graves2006connectionist} to supervise the training:
\begin{equation}
\begin{aligned}
LossLR = -\ln p(t \mid l),
\end{aligned}
\label{equ:loss_lip generation}
\end{equation}
where $p(t \mid l) = \sum _i {p(\pi _i \mid l)}$ is the probability of generating the target $t$, which is the sum over all possible CTC paths $\pi = \{ \pi _1, \pi _2, ... \}$. For each possible path $\pi _i$, the probability is $p(\pi _i \mid l) = \prod ^K_{k=1} {p(\hat{t}_k \mid l)}$ .

\subsection{Lip Generation}
\label{sec:lipg}
Given a text sequence $t$ and a guide lip image $I$, the lip generation model generates a sequence of lip frames $\hat{l} = \{\hat{l}_1, \hat{l}_2, ..., \hat{l}_K\}$ that contains the lip speaking the text $t$. The main challenge of this task is how to align text and video frames. A simple scenario is that we can get the number of corresponding video frames (duration) of each character in the text, which can be extracted from the audio by some tools such as Penn Phonetics Lab Forced Aligner (P2FA)~\cite{yuan2008speaker}. In this case, we design a model (lip generation model w/ duration) adopting the duration to align between text and video frames. For the general case without pre-given duration information, we propose another model (lip generation model w/o duration) to learn the alignment between text and video frames through attention mechanism~\cite{bahdanau2014neural}.

\subsubsection{Lip Generation Model w/ Duration}
\label{sec:lg_dur}

Figure \ref{fig:lip_generation_model}(a) shows the structure of lip generation model w/ duration. The original text $t$ together with the duration vector $d$ is fed into an $Expander$ to get the unrolled text $t^* = \{t^*_1, t^*_2, ..., t^*_K\}$, whose length $K$ is the same as the number of lip video frames $l = \{l_1, l_2, ..., l_K\}$. For example, given $t = \{ t_1, t_2, t_3 \}$, $d = \{ 1,2,3 \}$, we can get unrolled text $ t^* = \{ t_1, t_2, t_2, t_3, t_3, t_3 \}$. Then we feed $t^*$ into a text encoder $E_T$ to extract the text features $z^t = E_T(t^*) = \{z^t_1, z^t_2, ..., z^t_K\}$ . Meanwhile, a guide lip image $I$ random sampled from video frames $l$ is fed into an image encoder $E_I$ to extract the image feature $z^I = E_I(I)$ . Each text feature $z^t_k$ and the image feature $z^I$ are further concatenated to generate a hybrid feature $z_k = [z^I, z^t_k]$, where both the text and visual information are incorporated. Following~\cite{song2018talking}, the hybrid feature is later fed into an RNN to enforce the temporal coherence. Finally taking the output of RNN as input, we apply an image decoder $Dec$ to generate the target frame $\hat{l_k} = Dec(RNN(z_k))$. We adopt skip connections between the image encoder $E_I$ and image decoder $Dec$ to preserve the fine-grained input features. An $L_1$ loss is used to supervise the training of lip generation, which tends to reduce blurring compared with $L_2$ Loss~\cite{isola2017image}.

\begin{equation}
\begin{aligned}
LossLG = \left \| \hat{l} _k - l_k \right \| _{1}
\end{aligned}
\label{equ:loss_lip generation}
\end{equation}

\subsubsection{Lip Generation Model w/o Duration}
\label{sec:lg_nodur}

The structure of the lip generation model w/o duration is presented in Figure \ref{fig:lip_generation_model}(b). There are two main changes compared with Figure \ref{fig:lip_generation_model}(a): 1) One is removing the $Expander$ in the input side. Instead, we directly use the original text as input. 2) The other is adding the attention module between encoder and decoder. Unlike in the case of w/ duration, where there is a corresponding input character when predicting each frame, now the network must learn the alignments between text and video frames. Attention mechanism is demonstrated to be capable of learning alignment in tasks such as neural machine translation~\cite{bahdanau2014neural} and text to speech~\cite{shen2018natural}. Considering the nature of the monotonic alignment between input text and output video frames, we adopt the location sensitive attention~\cite{chorowski2015attention}, which extends the content-based attention to focus on location information and encourages the alignment to move forward consistently. Moreover, the decoder is trained in an autoregressive manner, i.e., the output of the last frame will be used as input to predict the next frame.

In the next subsection, we introduce the talking face generation system based on the lip generation model in DualLip.

\subsection{Text to Talking Face Generation System}
\label{sec:ttf}
\begin{figure}[htb]
  \centering
  \includegraphics[width=\linewidth]{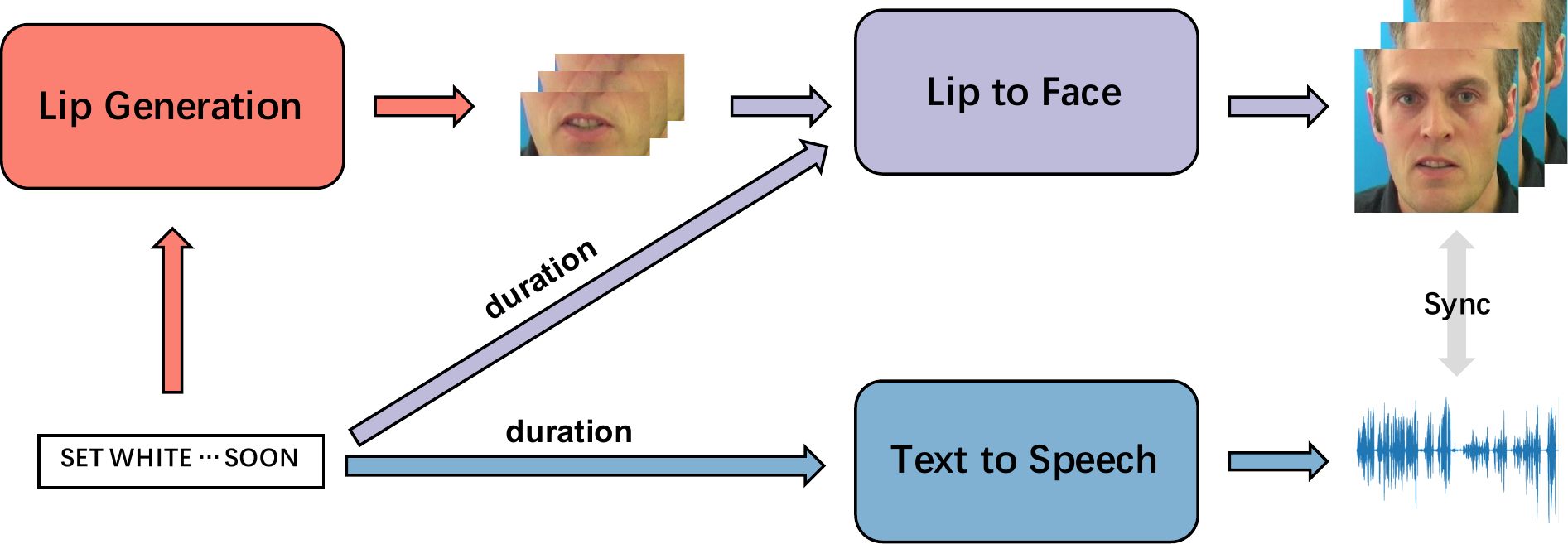}
  \caption{The pipeline of text to talking face generation system.}
  \label{fig:talking_face_overview}
\end{figure}

In order to leverage the benefits of DualLip, we extend the lip generation model of DualLip into a talking face generation system by introducing two additional modules: 1) lip to face (LTF), which is used to generate the full talking face given generated talking lip, and 2) text to speech (TTS), which generates the speech corresponding to the talking face. Different from the existing talking face generation systems that either generate talking face directly from speech, or generate speech from text first and then generate talking face from speech in a serial manner, our system generates in a parallel manner, which means speech and talking face are generated from text at the same time. The main challenge of this manner is the synchronization between video and speech. We propose to synchronize through duration and introduce an LTF model and a TTS model that both support customized duration. The pipeline of our talking face generation system is presented in Figure \ref{fig:talking_face_overview}.

\subsubsection{Lip to Face Generation}
\label{sec:ltf}

\begin{figure}[htb]
  \centering
  \includegraphics[width=\linewidth]{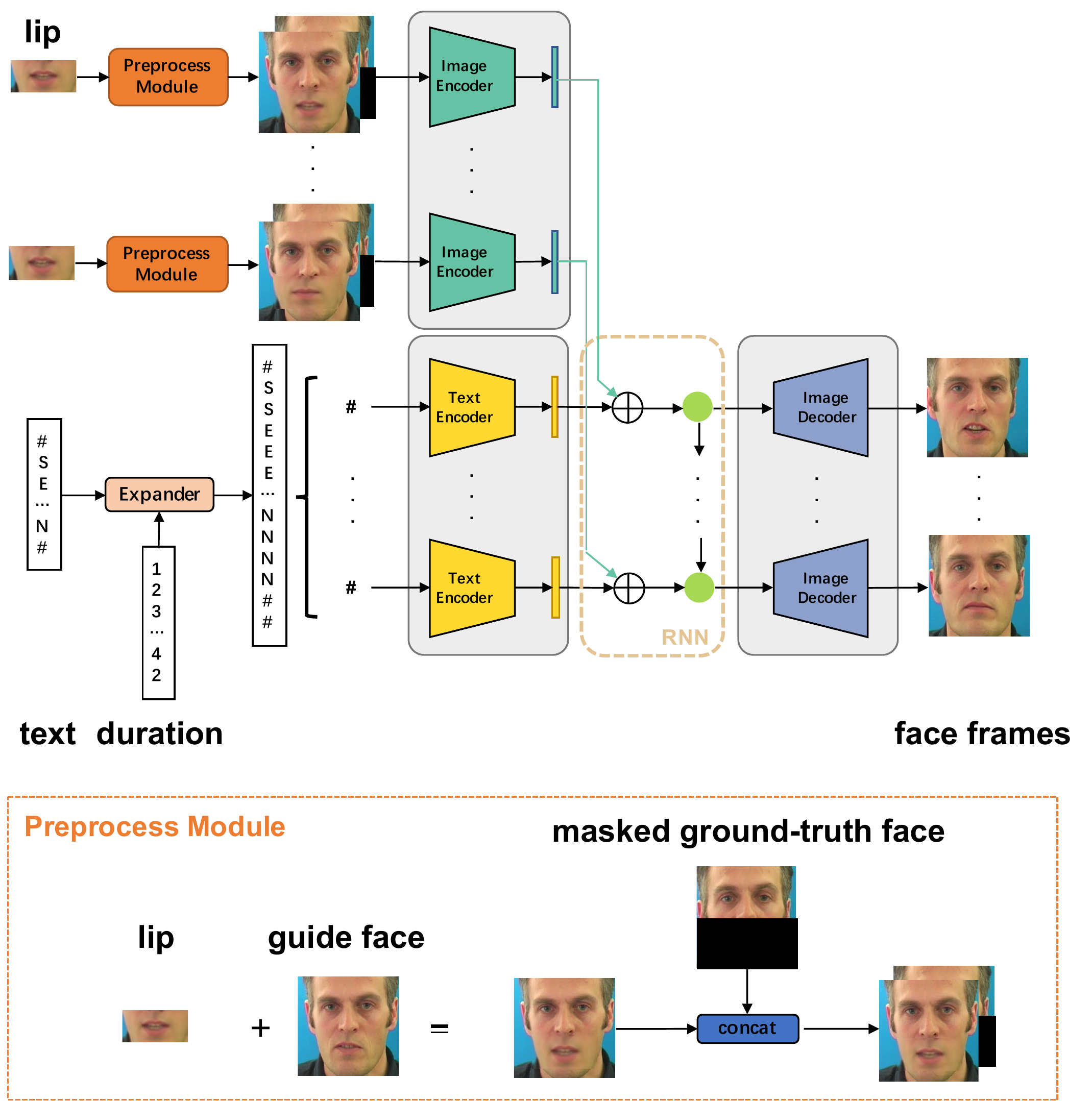}
  \caption{The structure of lip to face generation model.}
  \label{fig:ltf_model}
\end{figure}

As shown in Figure \ref{fig:ltf_model}, the structure of LTF model is similar to the lip generation model w/ duration, between which the main difference is the input of the image encoder. The input of LTF is a sequence of synthesized faces $F=\{F_1, F_2, ..., F_K\}$, which are generated by the Preprocess Module. The Preprocess Module first superimposes the lip generated by the lip generation model on a guide face, and then concatenates it channel-wise with the masked ground-truth face, which is generated by masking the lower half of the ground-truth face image. 
The masked ground-truth face provides information of the target pose while ensuring the network never gets any information of the ground-truth lip shape~\cite{kr2019towards}. Therefore, the input size of the image encoder is $H\times W \times 6$. In the inference stage, different frames use the same masked ground-truth face, which can avoid inconsistency caused by different face sizes in different frames.
The unrolled text $t^*$ is fed into the text encoder $E_T$ to get the text features $z^t = E_T(t^*) = \{z^t_1, z^t_2, ..., z^t_K\}$. For the $k_{th}$ frame of synthesized faces, the image encoder $E_I$ extracts the image feature $z^F_k = E_I(F_k)$. Then the concatenated feature $z_k = [z^F_k, z^t_k]$ are fed into an RNN and image decoder $Dec$ to get the $k_{th}$ generated face frame $\hat{F}_k = Dec(RNN(z_k))$. The training procedure is supervised by the same $L_1$ loss as used in lip generation.

\subsubsection{Text to Speech}
\label{sec:tts}
We adopt FastSpeech~\cite{ren2019fastspeech} and its multi-speaker version~\cite{chen2020multispeech} as our TTS model, which is a feed-forward network based on Transformer~\cite{vaswani2017attention} to generate mel-spectrogram in parallel. It can accept phonemes/characters and the corresponding duration as input, and generate speech that conforms to the duration. This feature ensures synchronization between the face video generated by LTF and the speech generated by TTS.

\subsubsection{Synchronization Between Speech and Face using Duration.}
\label{sec:sync}
As shown in Figure \ref{fig:talking_face_overview}, the same duration is fed into the LTF module and TTS module, and thus synchronization between face video and speech can be achieved. There are two sources of duration. In the w/ duration case, we use the pre-given duration. In the w/o duration case, we extract duration from the attention alignment $A^{T \times K}$ learned by the lip generation model, where $T$ is the length of input text and $K$ is the length of output video. Following~\cite{ren2019fastspeech}, the duration extractor is formed as $d_i = \sum^K_{k=1}{[\arg\max A_ {k} = i]}$, where $d_i$ is the duration of the $i_{th}$ character, and $A_k$ is the $k_{th}$ column of the attention alignment.

\begin{table*}
\centering
\begin{center}
\begin{tabular}{*{8}{c}}
\toprule
\multirow{2}{*}{Setting}&
\multicolumn{2}{c}{Training Data}&
\multicolumn{2}{c}{Lip Reading}&
\multicolumn{3}{c}{Lip Generation}\cr
    & Paired   & Unpaired
    & CER/PER (\%) $\downarrow$ & WER (\%) $\downarrow$
    & PSNR $\uparrow$       & SSIM $\uparrow$         & LMD $\downarrow$ \cr
\midrule
\multirow{3}*{GRID (w/ duration)}
~                         & 10\%  & 0\%   & 10.9 & 24.8 & 29.87  & 0.881 & 1.374 \cr
~                         & 10\%  & 90\%  & 2.66 & 6.25 & 30.91  & 0.902 & 1.234 \cr
~                         & 100\% & 0\%   & 1.30 & 3.10 & 30.68  & 0.895 & 1.288 \cr
\hline
\multirow{3}*{GIRD (w/o duration)}
~                         & 10\%  & 0\%    & 10.9 & 24.8 & 28.95  & 0.862 & 1.856 \cr
~                         & 10\%  & 90\%   & 4.61 & 11.9 & 29.30  & 0.871 & 1.811 \cr
~                         & 100\%  & 0\%   & 1.30 & 3.10 & 29.28  & 0.869 & 1.850 \cr
\hline
\multirow{3}*{TCD-TIMIT}  & 10\%  & 0\%    & 76.1 & -    & 25.45  & 0.782 & 2.114 \cr
~                         & 10\%  & 90\%   & 69.8 & -    & 27.64  & 0.823 & 1.711 \cr
~                         & 100\%  & 0\%   & 46.2 & -    & 29.33  & 0.857 & 1.314 \cr
\bottomrule

\end{tabular}
\end{center}
\caption{The results of DualLip. Note that CER is used for GRID, while PER is used for TCD-TIMIT. }
\label{tab:dual_basic_results}
\end{table*}

\section{Experimental Setup}

\subsection{Datasets and Preprocessing}

\subsubsection{GRID}
\label{sec:grid}

The GRID~\cite{Cooke2006An} corpus contains 34 speakers, but only 33 speakers' videos are available with a total of 32669 sentences. The sentences are in a fixed 6-word structure with a restricted grammar: $command + color + preposition + letter + digit + adverb$, e.g., ``\textit{set white with p two soon}''. There are 51 words in total, including 4 commands, 4 colors, 4 prepositions, 25 letters, 10 digits and 4 adverbs, yielding 64,000 possible sentences. All videos are of the same length (75 frames) with a frame rate of 25fps. Following~\cite{Leyuan2019LipSound}, we randomly select 255 sentences from each speaker for evaluation, leading to speaker-dependent results.

\subsubsection{TCD-TIMIT}

TCD-TIMIT~\cite{Harte2015TCD} contains 59 speakers uttering approximately 100 phonetically rich sentences each. Both the video length and sentence length in TCD-TIMIT are longer than GRID and not fixed, which is closer to the natural scene and more challenging. We follow the recommended speaker-dependent train-test splits in~\cite{Harte2015TCD}.

\subsubsection{Preprocessing}

For the videos, we first extract 68 facial landmarks by Dlib~\cite{King2009Dlib}, and then apply affine transformation to get an aligned face with the resolution of $256 \times 256$ . A $160 \times 80$ mouth-centered region is further extracted from the aligned face and resized to $128 \times 64$ as the final lip. And the aligned face is resized to $128 \times 128$ as the ground truth of lip to face generation model. For the texts, we use a $<\#>$ token to represent silent frames for the lip generation model w/ duration, and use a $<\#>$ token and $<\$>$ token to represent start-of-sentence silent frames and end-of-sentence silent frames respectively for the lip generation model w/o duration. Especially, we use phoneme instead of text in the TCD-TIMIT dataset following previous works~\cite{Harte2015TCD}.

\subsection{Evaluation Metrics}

We use character error rate (CER) and word error rate (WER) to measure the performance of lip reading on the GRID dataset, and phoneme error rate (PER) on the TCD-TIMIT dataset since the output in this dataset is phoneme sequence. As for lip generation and talking face generation, we use standard reconstruction metrics PSNR and SSIM~\cite{Wang2004Image} to evaluate the image quality of generated videos. To further evaluate whether the lip movements are accurate, we adopt Landmark Distance (LMD)~\cite{Chen2018Lip} to calculate the Euclidean distance between each corresponding pairs of facial landmarks on the generated video and ground truth. Slightly different from the calculation of LMD in~\cite{Chen2018Lip}, the resolution of the face in this work is $256 \times 256$ instead of $128 \times 128$. 
As for the LMD of lip generation model, we use the synthesized face before concatenating masked ground-truth face described in Section \ref{sec:ltf} for evaluation.

\subsection{Implementation Details}

\subsubsection{Model Configuration}
\label{sec:model}

As for the lip reading model, we adopt LipNet~\cite{Assael2016LipNet} and LipResNet~\cite{stafylakis2017combining} for GRID and TCD-TIMIT, respectively. The spatio-temporal visual module of LipNet is a stacked 3D CNN, while LipResNet combines 3D CNN with a ResNet18~\cite{He2016Deep} as the spatio-temporal visual module. As for the sequence processing module, the two models share the same structure that consists of a 2-layer bidirectional GRU and a linear layer.

\subsubsection{Training and Inference}

Our implementation is based on PyTorch~\cite{NEURIPS2019_9015}, and the networks are trained on two RTX 2080ti GPUs. We use AdamW~\cite{Loshchilov2017Decoupled} with an initial learning rate of 0.0002 and a decay ratio of 0.5 for lip reading, and Adam~\cite{Kingma2014Adam} with an initial learning rate of 0.001 and a decay ratio of 0.1 for lip generation and lip to face generation. The lip reading models are trained for $\sim$ 200 epochs, while the lip generation and lip to face generation models are trained for $\sim$ 30 epochs. In the DualLip training procedure, we first conduct supervised training until the models are close to convergence (before decaying the learning rate for the first time), and then add unsupervised training with the unsupervised loss coefficient $\alpha = 1$ for GRID and $\alpha = 0.1$ for TCD-TIMIT. For the guide image in lip generation and lip to face generation models, we randomly select one from the ground truth frames during training, and uniformly use the first frame in the inference phase. We train the FastSpeech model on the LibriTTS dataset~\cite{zen2019libritts}, and use WaveNet~\cite{oord2016wavenet} as the vocoder.

\section{Results}

In this section, we present experiment results to demonstrate our advantages in boosting the performance of lip reading, lip generation and talking face generation with unlabeled data, especially in low-resource scenarios with limited paired training data.

\subsection{Results of DualLip}

In order to verify the effectiveness of DualLip in improving the performance of lip reading and lip generation, we conduct a series of experiments on GRID and TCD-TIMIT. We first compare the results of these settings: 1) using only a small amount of paired data, 2) using a small amount of paired data and a large amount of unpaired data, 3) using the whole paired data. Then we fix the amount of paired data and change the amount of unpaired data to observe its effect on the performance. Finally, we train a lip reading model on the GRID dataset with all the paired data and additional unpaired data.

\begin{figure}[t]
  \centering
  \includegraphics[width=\linewidth]{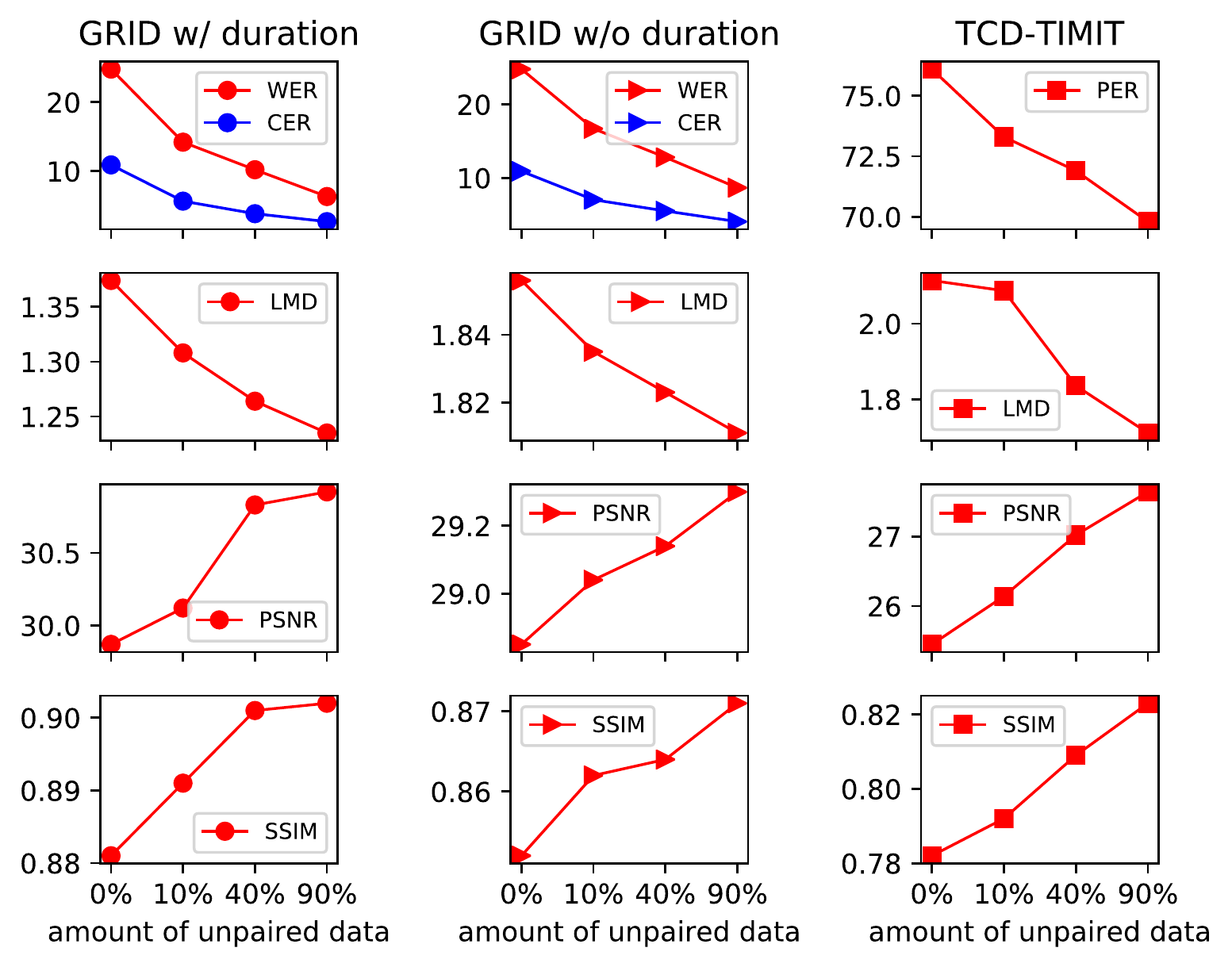}
  \caption{Results of varying the amount of unpaired data for DualLip.}
  \label{fig:dual_results}
\end{figure}

\subsubsection{Basic Results}
\label{sec:dual_basic_results}

On GRID, we conduct experiments in both w/ and w/o duration cases. We only use 10\% of the paired data, leaving the remaining 90\% as unpaired.\footnote{We conduct a comparative experiment to eliminate concerns about the implicit alignment signal in the unpaired data (since they are originally paired data). We randomly split the remaining 90\% data into two halves, each half consisting of lip-only and text-only data. Then we experiment with two settings of unpaired data combination to train DualLip: one is the first half of lip-only data and the second half of text-only data, and the other is the first half of lip-only and text-only data. The results show that there are no difference in the performance of lip reading and lip generation.}
On TCD-TIMIT, we experiment in the w/ duration case with 10\% paired data. According to the results presented in Table \ref{tab:dual_basic_results}, we can obtain the following observations:

1) Compared with the baseline trained with only a small amount of paired data, our method can greatly boost the performance of both lip reading and lip generation models by leveraging unpaired data with dual transformation.

2) Furthermore, using only 10\% paired data and 90\% unpaired data on GRID, our lip reading models achieve comparable results with those trained with the whole paired data, and our lip generation models even surpass them, which indicates the considerable advantage of our method in utilizing unlabeled data. During the training procedure of DualLip, the pseudo data pairs generated by the same unlabeled data are constantly changing. This is actually a kind of data augmentation, which explains why we can achieve better results than using the whole paired data.

3) Moreover, we observe that the lip reading performance in the w/ duration case is better than w/o duration. This is because the lip generation model w/ duration has better performance than the lip generation model w/o duration, resulting in better quality of generated lips, which further promotes the training of the lip reading model. The results on TCD-TIMIT also confirm this analysis that lip reading is more sensitive to the quality of generated data, while lip generation is more robust. The performance of the two baseline models on TCD-TIMIT is inferior, which means the low quality of generated data. However, we can see that lip generation can still benefit greatly from DualLip.

Above all, the results verify that DualLip can make full use of unpaired data to improve the performance of both lip reading and lip generation with limited training data.

\subsubsection{Varying Unpaired Data}

In order to observe the effect of the amount of unpaired data, we experiment by varying the amount of unpaired data while fixing the amount of paired data as 10\%. Figure \ref{fig:dual_results} shows the results. We can draw two observations: 1) As the amount of unpaired data increases, both lip generation and lip reading can achieve better performance. This is in line with our intuition and illustrates the importance of the amount of unpaired data under unsupervised learning. 2) As more unpaired data is added, although the performance growth becomes slower, the growth trend still continues, indicating the great potential of our method.

\subsubsection{Lip Reading SOTA on GRID}

We further conduct an experiment to explore the upper limit of DualLip's performance for lip reading on the GRID dataset. As mentioned in Section \ref{sec:grid}, the sentences in GRID are in a restricted grammar.
Therefore, we train DualLip with all the paired data and additional $\sim$40000 unpaired sentences, which are artificially synthesized according to the grammar.
Note that as the duration of the synthesized data is unknown, we train in the w/o duration case. As shown in Table \ref{tab:sota}, the lip reading model trained in this way not only exceeds the baseline trained with the whole paired data, but also obtains a new state-of-the-art performance on the GRID dataset.

\begin{table}
\begin{threeparttable}[b]
  \begin{tabular}{cccl}
    \toprule
    Method & CER (\%) $\downarrow$ & WER (\%) $\downarrow$ \\
    \midrule
    LipNet~\cite{Assael2016LipNet} & 1.9    & 4.8 \\
    LCANet~\cite{xu2018lcanet}     & 1.3    & 2.9 \\
    LipSound~\cite{Leyuan2019LipSound}    & 1.532    & 4.215 \\
    \hline
    LipNet (reproduce)   & 1.30    & 3.10 \\
    \textbf{DualLip}  & \textbf{1.16}   & \textbf{2.71} \\
    \bottomrule
  \end{tabular}
\end{threeparttable}
\caption{Our lip reading model based on DualLip achieves SOTA performance on GRID. Note that for fairness, the result of LipSound~\cite{Leyuan2019LipSound} is trained without external paired speech data. }
\label{tab:sota}
\end{table}

\subsection{Results of Talking Face Generation}

\subsubsection{Quantitative Evaluation}

We evaluate our text to talking face generation system quantitatively on GRID. The training data consists of 10\% paired data and 90\% unpaired data. For comparison, we implement the speech to face generation algorithm described in~\cite{song2018talking}. The structure of speech to face model is similar to the text to lip model w/ duration, except the input and the audio encoder instead of text encoder. The input for speech to face model is a guide face image and the 80-band mel-spectrogram, which is extracted from each audio segment ($T = 350ms$) with a hop size of 12.5ms and window size of 50ms following~\cite{kr2019towards}. The speech to face model is trained on 10\% paired data, which is consistent with our lip generation and lip to face model.
We can draw two conclusions from the results presented in Table \ref{tab:ttf}: 1) Compared with speech to face, the performance of using text as input in the w/ duration case is better than using audio as input. This is because audio contains a lot of redundant information compared to text, such as tone and noise, which will increase the burden of the network to learn a good feature.
2) DualLip can further take advantage of unlabeled data to improve the performance of talking face generation, when the paired training data is limited.

\begin{table}
\begin{threeparttable}[b]
  \begin{tabular}{ccccl}
    \toprule
    Setting                                      & PSNR $\uparrow$   & SSIM $\uparrow$ & LMD $\downarrow$ \\
    \midrule
    speech to face~\cite{song2018talking}        & 27.61  & 0.858 & 1.586 \\
    \hline
    lip generation\tnote{1} + lip to face                 & 27.77  & 0.862 & 1.417 \\
    lip generation\tnote{1} + lip to face + DualLip       & 27.90  & 0.866 & 1.323 \\
    \hline
    lip generation\tnote{2} + lip to face             & 27.21  & 0.841 & 2.002 \\
    lip generation\tnote{2} + lip to face + DualLip   & 27.30  & 0.852 & 1.920 \\
  \bottomrule
\end{tabular}
\begin{tablenotes}
  \item [1] lip generation w/ duration.
  \item [2] lip generation w/o duration.
\end{tablenotes}
\end{threeparttable}
\caption{Quantitative results of talking face generation.}
\label{tab:ttf}
\end{table}

\subsubsection{Qualitative Evaluation}

\begin{figure}[htb]
  \centering
  \includegraphics[width=\linewidth]{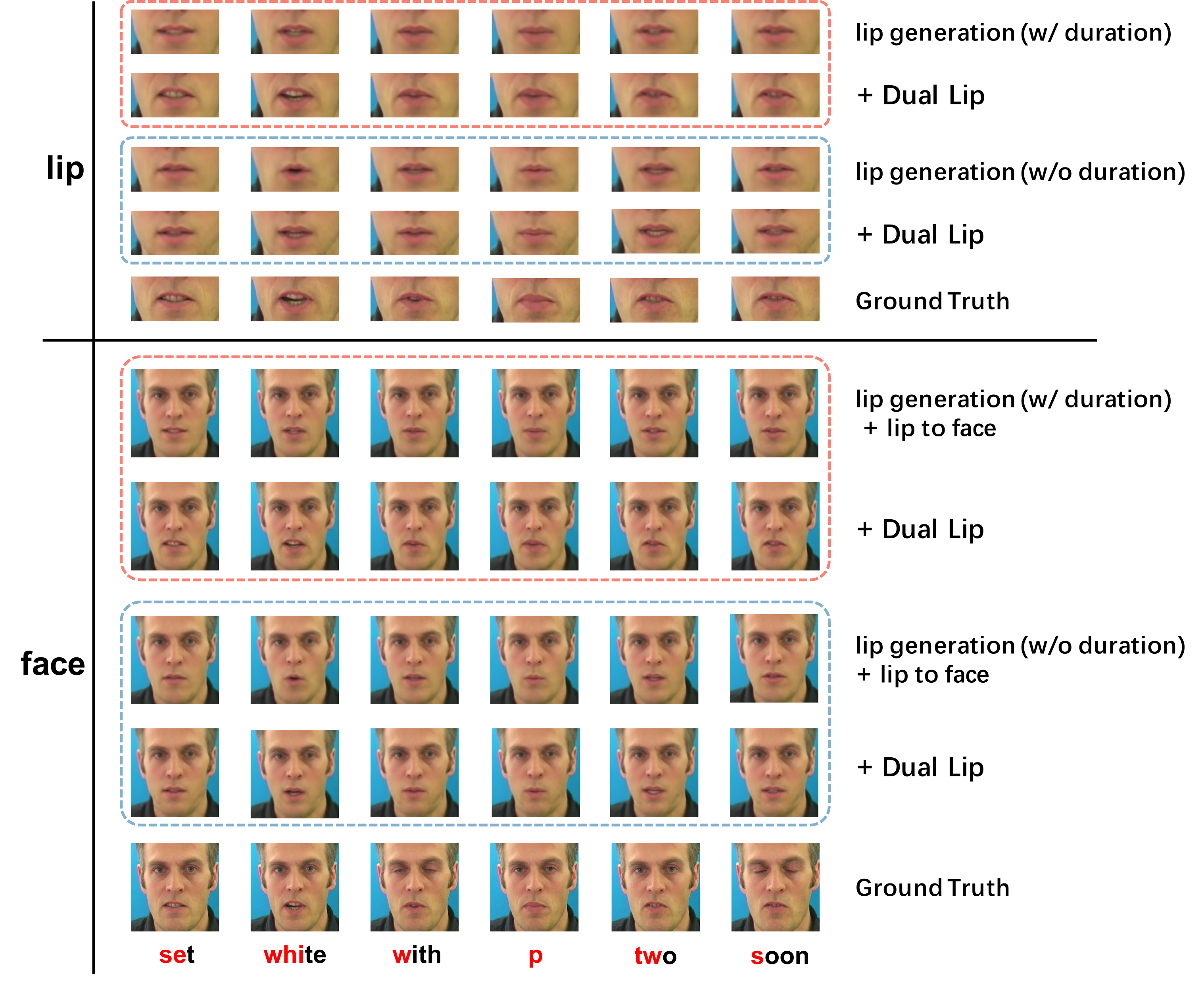}
  \caption{Visualization for talking face generation. Each column comes from the same frame of the generated videos when they are trying to speak specific segments of the words (red part) shown in the last row. Zoom in for better resolution.}
  \label{fig:talking_face_demo}
\end{figure}

Figure \ref{fig:talking_face_demo} shows the visualization results of talking face generation. Before using DualLip, the generated images are blurred in the lip area (especially the teeth), and the shape of the lips is not obvious. After using DualLip, we can see that the teeth are clearer, and the shape of the lips is highly close to the ground truth. The visual performance of the w/ duration case is better than w/o duration, which is consistent with the results in the quantitative evaluation.
More video demos can be found in our project page \url{https://dual-lip.github.io}.

\subsubsection{Alignment} 
\begin{figure}[htb]
  \centering
  \includegraphics[width=\linewidth]{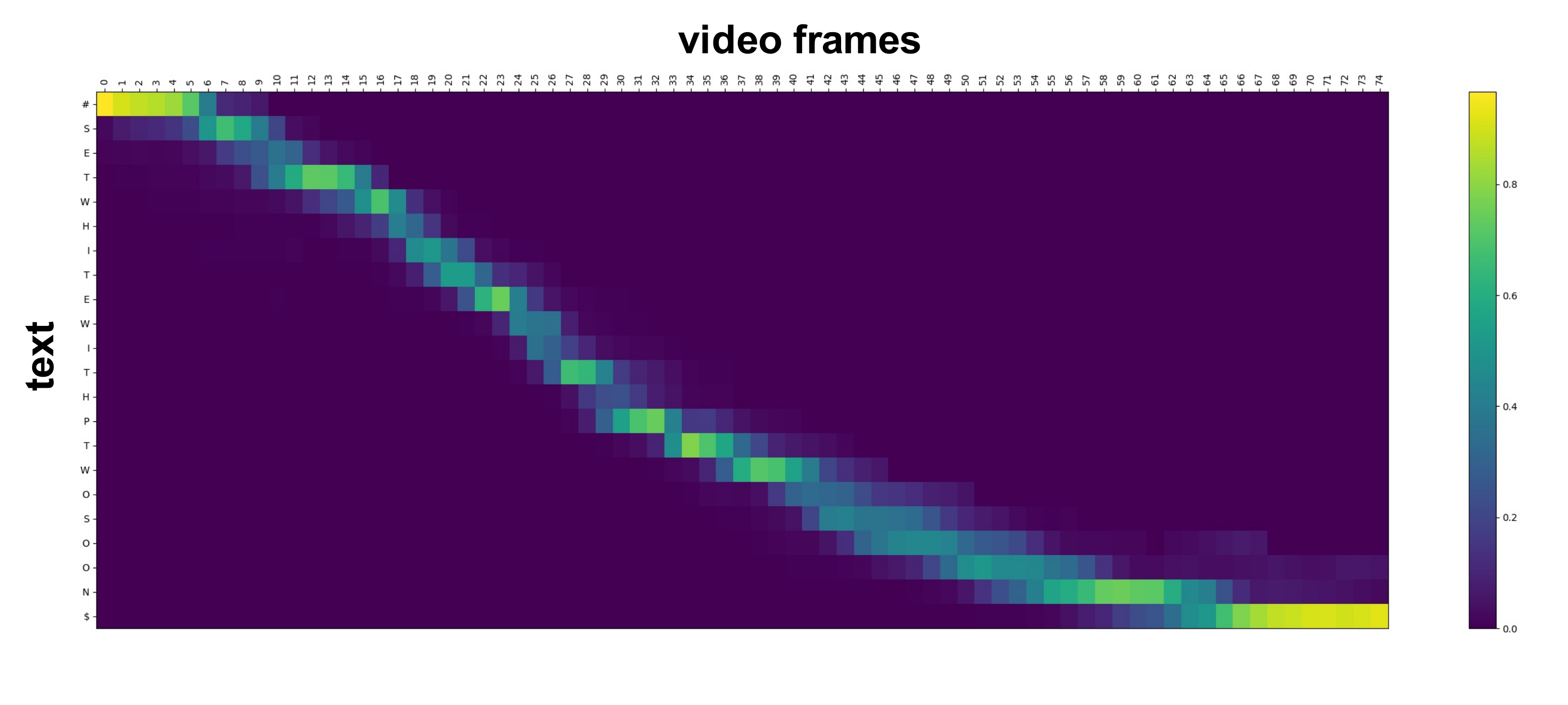}
  \caption{The alignments between text and image in each frame learned by lip generation model w/o duration.}
  \label{fig:attn}
\end{figure}

Figure \ref{fig:attn} presents the alignment learned by lip generation model w/o duration, which is smooth and monotonous. We can find that the short-pronounced characters (e.g., ``e'' in the word ``set'') have fewer continuous frames in the alignment, which is consistent with the duration. Therefore, we can extract the duration through the learned alignment and achieve accurate synchronization between talking face and speech.

\section{Summary and Discussions}

In this work, we developed the DualLip system to jointly improve lip reading and generation by leveraging their task duality, and improve their models trained on limited paired text and lip video data with the help of extra unlabeled data. The key idea is to perform dual transformation between unlabeled text and lip video data, i.e., transforming unlabeled text to lip video using a lip generation model to form  pseudo text and lip video pairs, and training the lip reading model with the pseudo pairs, and vice versa. Experiment results on GRID and TCD-TIMIT datasets demonstrate that our systems can effectively utilize unlabeled data to improve the performance of lip reading, lip generation and talking face generation. Specifically, on GRID, our lip generation model trained with only 10\% paired data together with 90\% unlabeled data exceeds the performance of using all paired data, and our lip reading model outperforms the previous state-of-the-art result.

Our DualLip and text to talking face generation systems have a number of possible applications and we will explore in the future. 1) Our DualLip system can take advantage of the widespread unlabeled video and text data to improve the performance of lip reading, which is beneficial to speech recognition especially in noisy environments. 2) Video conferencing becomes more and more popular, which poses a great challenge to the bandwidth of network transmission. With the help of our text to talking face generation system, only text needs to be transmitted over network, and the corresponding talking face video can be generated locally, which significantly reduces network traffic. 3) Current virtual assistants such as Siri and Alexa can only make voices, but no corresponding videos. Our text to talking face generation system can enhance those virtual assistants with generated talking face videos. Our system can also be used to create virtual characters and cartoons.





\begin{acks}

This work was supported by the \grantsponsor{GS501100001809}{National Natural Science Foundation of China}{https://doi.org/10.13039/501100001809} under Grant No.:~\grantnum{GS501100001809}{61832007},~\grantnum{GS501100001809}{61622403},~\grantnum{GS501100001809}{61621091},~\grantnum{GS501100001809}{U19B2019}.
\end{acks}

\bibliographystyle{ACM-Reference-Format}
\bibliography{acmart}

\end{document}